\documentclass[prc,aps,manuscript]{revtex4}
\usepackage{graphicx}
\usepackage{dcolumn}
\usepackage{bm}

\newcommand{\bs}{\bigskip}
\newcommand{\bc}{\begin{center}}
\newcommand{\ec}{\end{center}}

\begin{document}

\title{Non-Markovian large amplitude motion and nuclear fission}
\author{ V. M. \surname{Kolomietz} 
\thanks{Electronic address: vkolom@kinr.kiev.ua}}
\author{ S. V. \surname{Radionov} 
\thanks{Electronic address: Sergey.Radionov@matfys.lth.se}}
\affiliation{\textit{Institute for Nuclear Research, 03680 Kiev, Ukraine} }
\date{\today}

\begin{abstract}
The general problem of dissipation in macroscopic large-amplitude collective
motion and its relation to energy diffusion of intrinsic degrees of freedom
of a nucleus is studied. By applying the cranking approach to the nuclear
many body system, a set of coupled dynamical equations for the collective
classical variables and the quantum mechanical occupancies of the intrinsic
nuclear states is derived. Different dynamical regimes of the intrinsic
nuclear motion and its consequences on time properties of collective
dissipation are discussed. The approach is applied to the descant of the
nucleus from the fission barrier.
\end{abstract}


\maketitle

\section{Introduction}

Nuclear large scale dynamics (nuclear fission, heavy ion collisions etc.) is
a good probe for the investigation of complex time evolution of finite Fermi
systems. The conceptual question is here how collective modes of motion
appear in a system with many degrees of freedom and how they interact with
all other intrinsic modes. Nuclear collective dynamics can be studied by
using the concept of macroscopic motion for a few collective degrees of
freedom, which are chosen to describe gross properties of the nucleus 
\cite%
{sije87,hamy88,hofm97}. Such a kind of approach is acceptable for a slow
collective motion where the fast intrinsic degrees of freedom exert forces
on the collective variables leading to a transport equation. The crucial
point of such an approach is the separation of the total energy of the
system into potential energy, collective kinetic energy and dissipation
energy. Moreover the dissipation of collective motion implies fluctuations
in the corresponding collective variables, as follows from the
fluctuation-dissipation theorem \cite{bale75}.

Dissipation of the nuclear collective energy reveals itself, for instance,
as the non-zero contribution of the internucleonic collisions to the widths
of the nuclear giant multipole resonances. On the other hand, the
experimental observation of the finite variance of the kinetic energy of the
fission fragments manifests the fact that fluctuations have to be also
associated with the collective variables. Both the dissipation and the
fluctuations can be described by the introduction of friction and random
forces, related to each other by the fluctuation-dissipation theorem. In
this respect, the Fokker-Planck or Langevin approaches can be used to study
the nuclear large scale dynamics, see Refs. \cite{bosu93,abay96} and
references therein. In general, basic equations of motion for the
macroscopic parameters, describing complex dynamics of the many-body systems
like nuclei, have non-Markovian structure. One of the first considerations
of memory (non-Markovian) effects for classical liquids can be found in Ref.
\cite{fren46}. For the dynamics of nuclei, memory effects have been
investigated within different approaches. In this respect, one can mention
the dissipative diabatic model \cite{ayno82}, the linear
response theory \cite{hofm97} and the fluid dynamic approach 
\cite{kosh04,kora01}. In this paper, we would like to apply the
non-Markovian dynamics to the study of the nuclear fission's characteristics
and clarify the role of the fluctuation and memory effects in the descant of
the nucleus from the fission barrier to the scission point.

The plan of the paper is as follows. In Sect.~II we derive the non--Markovian
Langevin equation of motion for the nuclear shape variables starting from
the collisional kinetic equation. Sect. III is devoted to details of the
numerical determination of the saddle--to--scission time and pre--scission
kinetic energy in presence of the memory effects and the random force for
the descent of the nucleus from the barrier to the scission point. Summary
and conclusions are given in Sect.~IV. 
We assume that dynamics of nuclear many body system can be described as a
coupled motion of macroscopic collective modes and intrinsic nucleonic ones.
The slow collective modes of the nuclear large--amplitude motion are treated
in terms of a set of classical time--dependent variables $\vec{q}(t)\equiv
\{q_{1}(t),q_{2}(t),...,q_{N}(t)\}$, while the fast intrinsic modes are
described quantum mechanically through the time--evolution of occupancies of
nucleonic many body states.

The intrinsic dynamics can be determined through the Liouville equation for
the density matrix operator $\hat{\rho}$,
\begin{equation}
\frac{\partial \hat{\rho}(t)}{\partial t}+i\hat{L}(t)\hat{\rho}(t)=0,
\label{L}
\end{equation}
where $\hat{L}$ is the Liouville operator defined in terms of the commutator,
$\hat{L}\hat{\rho}=\left[ \hat{H},\hat{\rho}\right]/\hbar$ of the nuclear 
many body Hamiltonian $\hat{H}(\vec{q}[t])$.

Using Zwanzig's projection technique~\cite{zw} and a basis of adiabatic eigenenergies
$E_k$ and eigenfunctions $\Psi_k$ of the nuclear many body Hamiltonian $\hat{H}(\vec{q}[t])$
\cite{kar}, we can get dynamical equations for a non--diagonal part of the density
matrix,
\begin{equation}
\rho_{nm}(t)=-i\sum_i \int_0^t dt'\dot{q}_i(t')\frac{{\rm exp}[-i\omega_{nm}(t-t')]}
{\omega_{nm}}\left[h_{i,mn}(t')\rho_{nn}(t')-h_{i,nm}(t')\rho_{mm}(t')\right],
\label{rnm}
\end{equation}
and its diagonal part,
\begin{eqnarray}
\frac{\partial \rho_{nn}(t)}{dt}=\frac{2}{\hbar^2}\sum_{i,j}\dot{q}_i(t)
\int_0^t dt'\dot{q}_j(t')\sum_{m\neq n}h_{i,nm}(t)h_{j,mn}(t')
\frac{{\rm cos}[\omega_{nm}(t-t')]}{\omega_{nm}^2}
\nonumber\\
\times\left[\rho_{mm}(t')-\rho_{nn}(t')\right].
\label{rnn}
\end{eqnarray} 
Here, $\omega_{nm}=(E_n-E_m)/\hbar$ and matrix elements 
$h_{i,nm}=\big|\partial \hat{H}/\partial q_i \big|_{nm}$ measure the coupling between
the quantum nucleonic and the macroscopic collective subsystems.

To proceed further, we use a random matrix theory developed in our previous paper~\cite{kar} 
for the case of a single collective coordinate. We omit all intermediate steps and give a
basic diffusion--like equation of motion for the ensmble averaged occupancies 
$\bar{\rho}(E,t)$ of the many body states $E\equiv E_n$:
\begin{eqnarray}
\Omega(E)\frac{\partial \bar{\rho}(E,t)}{\partial t}=\sum_{i,j}s_{ij}
\dot{q}_i(t)\int_0^t dt'{\rm exp}\left(-\frac{|t-t'|}{\hbar/\Gamma_{ij}}\right)
Y_{ij}(\vec{q}[t],\vec{q}[t'])\dot{q}_j(t')
\nonumber\\
\times\frac{\partial }{\partial E}
\big[\Omega(E)\frac{\partial \bar{\rho}(E,t')}{\partial E}\big],
\label{reqm}
\end{eqnarray}
where $s_{ij}$ and $\Gamma_{ij}$ are, correspondingly, the strengths and widths 
of the energy distributions of the ensemble averaged coupling matrix elements 
$h_{i,nm}$, $Y(\vec{q}(t),\vec{q}(t')$ are the correlation functions measuring 
how strong the coupling matrix elements correlate at different collective deformation 
parameters $\vec{q}(t)$ and $\vec{q}(t')$ , and $\Omega(E)$ is the nuclear many body 
level density at excitation energy $E$.

\section{Energy rate}

In order to define properly dynamics of the classical collective parameters $\vec{q}(t)$ 
within the cranking approach, one has to consider total energy of the nuclear many body system, 
which can be written as
\begin{equation}
\Sigma (t)=E_{\mathrm{gs}}(q)+\mathrm{Tr}\{\hat{H}[\vec{q}(t)]\ \hat{\rho}(t)\}.
\label{Etot}
\end{equation}
Differentiating over time both sides of Eq.~(\ref{Etot}), we get
\begin{equation}
\frac{d\Sigma }{dt}=\sum_{i}\dot{q}_{i}\frac{\partial E_{\mathrm{gs}}}
{%
\partial q_{i}}+\sum_{i}\dot{q}_{i}\sum_{n,m}\left( \frac{\partial \hat{H}}
{%
\partial q_{i}}\right) _{mn}\rho _{nm}+\sum_{n}E_{n}\frac{\partial \rho _{nn}}{\partial t}
+\sum_{i}\dot{q}_{i}\sum_{n}\left( \frac{\partial \hat{H}}{\partial q_{i}}\right) _{nn}\rho _{nn}.  
\label{dEdt}
\end{equation}
The first term in the right--hand side of~\ref{dEdt}) describes a change of the collective potential energy. 
The second term depends on the non–-diagonal component of the density matrix $\rho_{nm}(t)$ . Its time evolution is 
caused by the virtual transitions among adiabatic states of the nuclear many body Hamiltonian. We believe that 
such a term is a microscopic source for the appearance of the collective kinetic energy:
\begin{equation}
\left(\frac{d\Sigma}{dt}\right)^{virt}\approx \sum_i\dot{q}_i\sum_j B_{ij}(\vec{q})\ddot{q}_j,
\label{Evirt}
\end{equation}
where $B_{ij}$ is a collective mass parameter,
\begin{equation}
B_{ij}=\sum_{n,m}h_{i,nm}h_{j,mn}\omega_{nm}^{-3}[\rho_{mn}-\rho_{nn}].
\label{Bij}
\end{equation}
The third term in the right--hand side of~\ref{dEdt}) is defined by the real transitions between
adiabatic many body states thus, defining dissipation of energy associated with the nuclear
collective motion:
\begin{eqnarray}
\left(\frac{d\Sigma}{dt}\right)^{real}=\sum_i\dot{q}_i(t)s_{ij}\int_{E_{gs}}^{+\infty}
dE E\Omega(E)\sum_j \int_0^t dt'{\rm exp}\left(-\frac{|t-t'|}{\hbar/\Gamma_{ij}}\right)
Y_{ij}(\vec{q}[t],\vec{q}[t'])\dot{q}_j(t')
\nonumber\\
\times\frac{\partial }{\partial E}
\left[\Omega(E)\frac{\partial \bar{\rho}(E,t')}{\partial E}\right],
\label{Ereal}
\end{eqnarray}
see Eq.~(\ref{reqm}). It can be shown that the fourth term in the right--hand side of~\ref{dEdt})
is described by the slopes of the adiabatic eigenenergies $\partial E_n/\partial q_i$
and within our random matrix approach is neglected compared to the other terms.

Collecting Eqs.~(\ref{Evirt}) and~(\ref{Ereal}), one obtains for (\ref{dEdt}),
\begin{equation}
\left(\frac{d\Sigma}{dt}\right)=\sum_i\dot{q}_i(t)F_i(\vec{q}[t],\dot{\vec{q}}[t],t),
\label{qF}
\end{equation}
where quantities $F_i$ mean forces acting on the collective subsystem given by 
\begin{eqnarray}
F_i=\sum_jB_{ij}\ddot{q}_j+\frac{\partial E_{gs}}{\partial q_i}+
\int_{E_{gs}}^{+\infty}
dE E\Omega(E)\sum_j \int_0^t dt'{\rm exp}\left(-\frac{|t-t'|}{\hbar/\Gamma_{ij}}\right)
Y_{ij}(\vec{q}(t),\vec{q}(t')\dot{q}_j(t')
\nonumber\\
\times\frac{\partial }{\partial E}
\left[\Omega(E)\frac{\partial \bar{\rho}(E,t')}{\partial E}\right].
\label{Fi}
\end{eqnarray}

\section{Equations of motion for the collective parameters}
\label{emcp}

To get equations of motion for unknown collective parameters, 
we assume that all partial contributions to the energy rate 
(\ref{qF}) are zero~\cite{sije87}
\begin{equation}
F_i\equiv 0.
\label{Fi0}
\end{equation}
Thus, collective dynamics satisfy the following set of equations 
\begin{eqnarray}
\sum_j B_{ij}[\vec{q}]\ddot{q}_j=-\frac{\partial E_{gs}}{\partial q_i}
-s_{ij}\int_{E_{gs}}^{+\infty}
dE E\Omega(E)\sum_j \int_0^t dt'{\rm exp}\left(-\frac{|t-t'|}{\hbar/\Gamma_{ij}}\right)
Y_{ij}(\vec{q}(t),\vec{q}(t')\dot{q}_j(t')
\nonumber\\
\times\frac{\partial }{\partial E}
\left[\Omega(E)\frac{\partial \bar{\rho}(E,t')}{\partial E}\right].
\label{eqm1}
\end{eqnarray}
can demonstrate that for the constant–temperature level density,
$\Omega()=const \cdot {\rm exp}(E/T)$, where $T$ is the temperature of the nucleus, 
we obtain a closed set of equations of motion for the collective parameters:
\begin{eqnarray}
\sum_j B_{ij}[\vec{q}]\ddot{q}_j=-\frac{\partial E_{gs}}{\partial q_i}
-\frac{s_{ij}}{T}\int_{E_{gs}}^{+\infty}
dE E\Omega(E)\sum_j \int_0^t dt'{\rm exp}\left(-\frac{|t-t'|}{\hbar/\Gamma_{ij}}\right)
Y_{ij}(\vec{q}(t),\vec{q}(t')\dot{q}_j(t')
\nonumber\\
\times\frac{\partial }{\partial E}
\left[\Omega(E)\frac{\partial \bar{\rho}(E,t')}{\partial E}\right].
\label{eqm2}
\end{eqnarray}
It should be pointed out that Eq.~(\ref{eqm2}) describes the ensemble averaged collective 
dynamics, i. e., averaged over many different random realizations of the intrinsic nucleonic 
subsystem. In this way, of course, we loose information about the quantum fluctuations of 
the intrinsic degrees of freedom of the nuclear many body system which, in principle, 
may be important. In order to try to take them into account somehow, we introduce a 
phenomenological stochastic force term, $\xi_i(t)$, into the collective equations of 
motion (\ref{eqm2}) and requiring that the fluctuation–-dissipation theorem is hold,  
\begin{equation}
\langle \xi_i(t)\xi_j(t')\rangle = Ts_{ij}Y_{ij}(\vec{q}[t],\vec{q}[t'])
{\rm exp}\left(-\frac{|t-t'|}{\hbar/\Gamma_{ij}}\right).
\label{fdt}
\end{equation}
this, the collective dynamics gets a form of the non--Markovian Langevin equations of motion 
\begin{eqnarray}
\sum_j B_{ij}[\vec{q}]\ddot{q}_j=-\frac{\partial E_{gs}}{\partial q_i}
-\frac{s_{ij}}{T}\int_{E_{gs}}^{+\infty}
dE E\Omega(E)\sum_j \int_0^t dt'{\rm exp}\left(-\frac{|t-t'|}{\hbar/\Gamma_{ij}}\right)
Y_{ij}(\vec{q}(t),\vec{q}(t')\dot{q}_j(t')
\nonumber\\
\times\frac{\partial }{\partial E}
\left[\Omega(E)\frac{\partial \bar{\rho}(E,t')}{\partial E}\right]+\xi_i(t).
\label{eqm3}
\end{eqnarray}

\section{Numerical calculations}
\label{nc}

Now we turn to the numerical determination of nuclear fission's characteristics. 
We study the symmetric fission of highly excited heavy nuclei whose space shape 
may be obtained by rotation of some profile function $W^2(z)$ around $z$--axis.
It is considered a 2–-parametric family of the Lorentz shapes~\cite{hamy88}:
\begin{equation}
W^2(z)=(z^2-\zeta_0^2)(z^2+\zeta_2^2)/Q,
\label{W2}
\end{equation}
where the multiplier $Q=-\zeta_0^3(\zeta_0^2/5+\zeta_2^2)$ guarantees the conservation 
of the nuclear volume. Here, all quantities of the length dimension are expressed in 
units of the radius $R_0$ of the spherical equal–volume nucleus. The parameter $\zeta_0$
defines
an elongation of the figure, while the parameter $\zeta _{2}$ is responsible
for 
the neck of the figure. Thus, in the case of $\zeta _{2}=\infty $ we
have a spheroidal 
shape and for $\infty <\zeta _{2}<0$ the neck appears.

The adiabatic collective potential energy of deformation $E_{gs}$ were taken from
\cite{hamy88}. The equations of motion (\ref{eqm3}), (\ref{fdt}) were solved numerically 
with the help of the simplest Euler method with the initial 
conditions corresponding to the saddle–point deformation and the initial kinetic energy 
$E_{\rm kin}=1~{\rm MeV}$ (initial neck velocity $\dot{\zeta_2}=0$). The numerical 
calculations were performed for the symmetric fission of a nucleus $^{236}{\rm U}$ at
temperature $T=2~{\rm MeV}$. We define the scission line from the condition of the 
instability of the nuclear shape with respect to any variations of the neck radius:
\begin{equation}
\frac{\partial^2 E_{gs}}{\partial \rho_{\rm neck}^2}=0,
\label{neck}
\end{equation}
where $\rho_{\rm neck}=\zeta_2/\sqrt{\zeta_0(\zeta_0^2/5+\zeta_2^2)}$ is the neck radius.

We considered a time of the nuclear descent from the saddle point to the scission~(\ref{neck}).
The total number of $2\times 10^4$ trajectories $\zeta_0(t),\zeta_2(t)$, generated by the 
different random realizations of the random forces $\xi_i(t)$, were taken into consideration 
in order to define the dynamic path of the system (\ref{eqm3}), (\ref{fdt}). To simplify the 
problem and clarify the role of the random force in the non–Markovian dynamics of the system, 
we stopped running trajectories $\zeta_0(t),\zeta_2(t)$ when they cross the line 
$\rho_{\rm neck}(\zeta_0,\zeta_2)=\rho_{\rm neck}^{\rm Newt}$, where $\rho_{\rm neck}^{\rm Newt}$
is the neck radius' value determined from the Newtonian non–Markovian dynamics ( i. e., when 
the stochastic terms are absent in Eqs.~(\ref{eqm3}), (\ref{fdt})).

In Fig. 1, we show the Langevin (solid line) and Newtonian (dashed line) dynamic trajectories 
of the neck radius $\rho_{\rm neck}(t)$. Horizontal line in the figure gives the scission value 
of the neck radius $\rho_{\rm neck}^{\rm Newt}$ derived from the Newtonian calculation.

A histogram of the distribution $p$ of the saddle–-to–-scission times $t_{\rm sc}$ is given in Fig. 2. 
We found that the most probable and mean value of the descent time are significantly smaller than the 
Newtonian estimation of $t_{\rm sc}$ shown by a vertical line. In fact, the random force speeds up the 
process of descent from the fission barrier. Indeed, the action of a random force, in some sense, 
"shakes loose" the system giving rise to a smaller time of motion between two given points comparing 
to the corresponding time for the unperturbed of the system. It should be pointed out that this is 
hold both for Markovian Langevin dynamics (it may be demonstrated analytically for some quite simple 
models) and non–-Markovian Langevin dynamics. The last feature is demonstrated in Fig. 3, where the 
mean value  $\langle t_{\rm sc}\rangle$ of the descent time is plotted versus the memory time  
$\tau\equiv \hbar/\Gamma_{ij}$ for the Langevin (solid line) and Newtonian (dashed line) paths of 
the system.

As seen in Fig. 3, the difference between the two non–Markovian calculations of the mean 
saddle–-to–-scission time grows with the memory time. This fact may be explained by the correlation 
properties of the random forces in the non–Markovian Langevin equations, see Eq.~(\ref{eqm3}). As 
can be seen from Eq.~(\ref{fdt}), the increase of the memory time $\tau$ amplifies the correlation   
in the collective coordinates $\zeta_0(t),\zeta_2(t)$ at two subsequent moments of time $t$ and
$t+\Delta t$. As a  consequence of that, one might expect that a quite big change, occurred for 
the coordinates $\zeta_0(t),\zeta_2(t)$ at time $t$, will give rise to a sufficiently big change for 
$\zeta_0(t),\zeta_2(t)$ even at the next time step $t+\Delta t$. This tendency will be stronger as 
far as the memory time $\tau$ grows up. Therefore, in average the system will reach the scission 
faster as compared to the non–-Markovian motion of the system.

\section{Summary and Conclusions}

Within non-Markovian Langeven approach, we have demonstrated a consistent description of nuclear 
large amplitude dynamics, including the memory effects and the random force. We have averaged the 
intrinsic nucleonic dynamics (\ref{rnm}), (\ref{rnn}) over suitably chosen statistics of the 
randomly distributed coupling matrix elements and the energy spacings. Owing to this procedure, 
we have derived the diffusion–-like equation of motion for the smeared occupancies 
$\bar{\rho}(E,t)$ of the adiabatic many body states. Note that the obtained equation of motion for 
$\bar{\rho}(E,t)$ is the non–-Markovian one, where the memory effects depend on the width of the 
randomly distributed matrix elements. 

Applying the ensemble averaging procedure, we have also derived the collective mass parameter 
and the internal energy rate. Finally, we derive a set of coupled dynamic equations for the 
macroscopic variables which take into consideration the time variation of the occupancies of 
the intrinsic nuclear states. Following the fluctuation–dissipation theorem, we have 
incorporated also the relevant random force into the macroscopic equations of motion. 
The main contribution to the rate of dissipation energy is due to the jump probabilities 
leading to a rate of dissipation which depends essentially on the total energy of the nucleus. 
The final result shows that a time irreversible energy exchange between the collective and 
internal degrees of freedom is possible when the level density increases with energy. We have 
applied our approach to the description of the descant of the nucleus from the fission barrier. 
We show that the random force accelerate significantly the process of descent from the barrier 
for both the Markovian and non–Markovian Langevin dynamics. We have observed that the difference 
between the two non–Markovian calculations (see Fig. 3) of the mean saddle–-to-–scission time grows 
with the memory time. This fact may be explained by 
the correlation properties of the random force in the non–-Markovian Langevin equations.

\bc
Figure captions
\ec

\bs

Fig.~1. Typical Langevin (solid line) and Newtonian (dashed line) trajectories of the neck radius
$\rho_{\rm neck}$ of the system (\ref{eqm3}), (\ref{fdt}). Horizontal line is the value of 
neck radius at the scission (\ref{neck}) $\rho_{\rm neck}^{\rm Newt}$derived from the corresponding 
Newtonian dynamics.

\bs

Fig.~2. Histogram of a probability density $p$ of the descent times $t_{\rm sc}$
for the non–-Markovian Langevin dynamics (\ref{eqm3}), (\ref{fdt}) at the memory
time $\tau\equiv\hbar/\Gamma_{ij}=2\times 10^{-23}~{\rm s}$, when 
the size of memory effects in the system is quite small. The vertical line gives 
the Newtonian estimation for the descent time.

\bs

Fig.~3. The mean value of the descent time $\langle t_{\rm sc}\rangle$
for the non–-Markovian Langevin dynamics (\ref{eqm3}), (\ref{fdt}) versus the memory time 
$\tau$ is shown by solid line. The Newtonian calculation of the descent time is given by 
the dashed line.

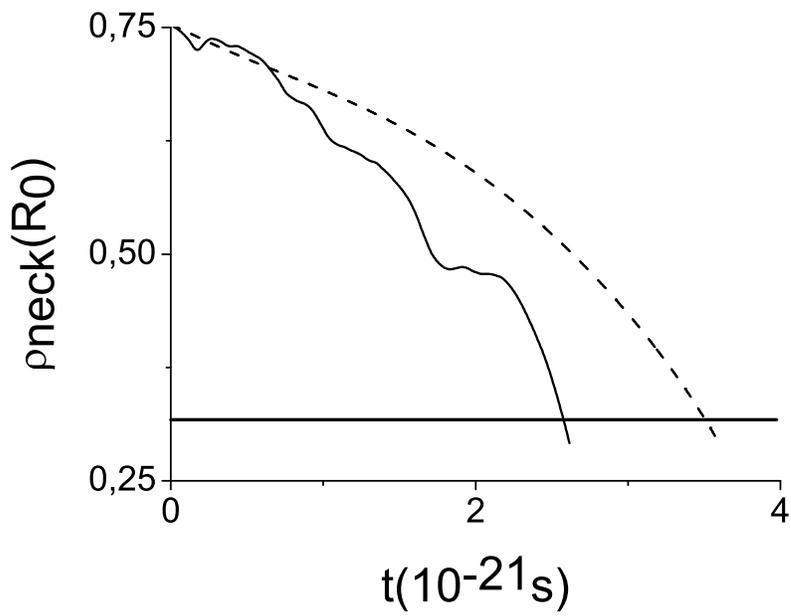
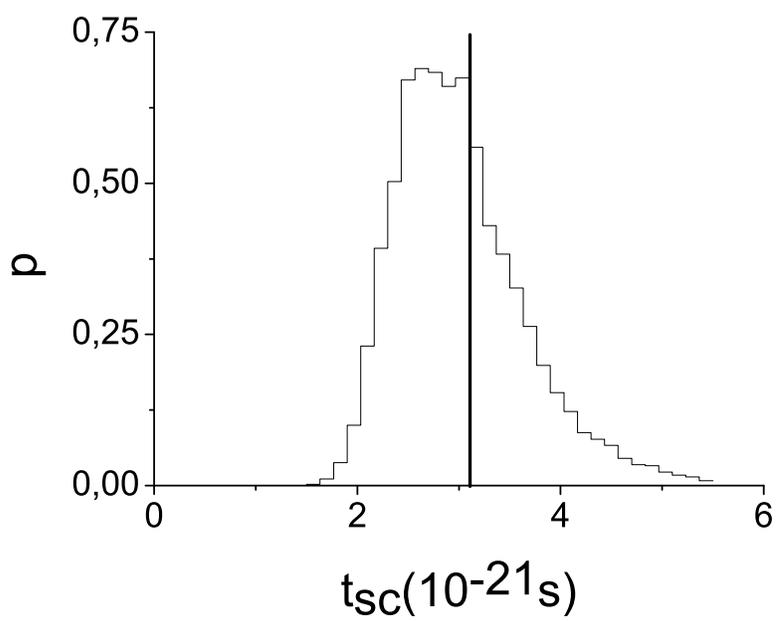
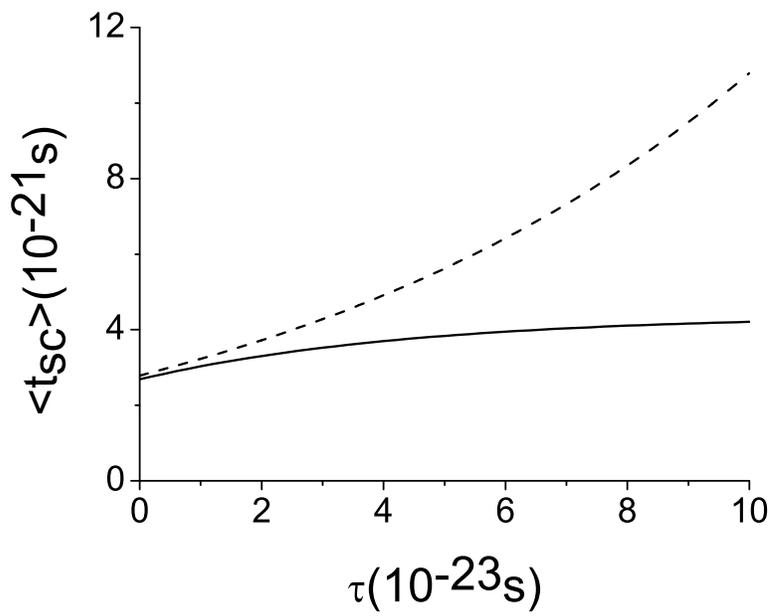
\end{document}